\documentclass[journal]{IEEEtran_RSIFSMA}
\usepackage[brazil]{babel}
\usepackage[T1]{fontenc} % para hifenização

\usepackage{adjustbox}
\usepackage{comment}
\usepackage{rotating}
\usepackage{realboxes}
\usepackage{makecell}

\usepackage[style=ieee]{biblatex}
\newcommand{\RSIFSMA}{Revista de Sistemas de Informação da FSMA}
\newcommand{\NUM}{8}
\newcommand{\ANO}{2011}
\newcommand{\PP}{2-7}
\newcommand{\HFP}{%
\begin{tabular}{p{\columnwidth}p{\columnwidth}}%
\hfill \RSIFSMA \hfill~                                                                       & \hfill \includegraphics[width=0.5\columnwidth]{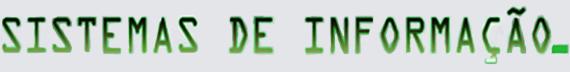} \hfill~  \\%
\hfill n. \NUM\ (\ANO) pp. \PP \hfill~                                               & \hfill http://www.fsma.edu.br/si/sistemas.html \hfill~ \\ \\%                                                                      
\includegraphics[width=\columnwidth]{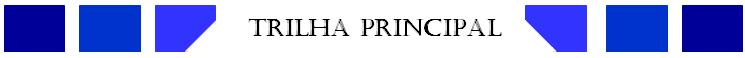} & \\%
\end{tabular}%
}
\newcommand{\TITULO}{Trends, Opportunities, and Challenges in Using Restricted Device Authentication in Fog Computing}
\ifCLASSINFOpdf
\else
\fi

\usepackage{lineno}
\usepackage{color, soul}

\usepackage{booktabs}
\usepackage{multirow}
\usepackage{subfigure}
\usepackage[brazil]{babel}

\begin{document}

%%%%%%%%%%%%%%%%%%%%%%%%%%%%%%%%%%%%%%%%%%%%%%%%%%%%%%%%%%%%%%%%%%%%%%%%%%%%
%*** NÃO ALTERE 
%%%%%%%%%%%%%%%%%%%%%%%%%%%%%%%%%%%%%%%%%%%%%%%%%%%%%%%%%%%%%%%%%%%%%%%%%%%%
%*** SE NECESSÁRIO, ESSAS INFORMAÇÕES 
%*** SERÃO ATUALIZADAS PELOS EDITORES
\renewcommand{\contentsname}{Sumário}
\renewcommand{\listfigurename}{Lista de Figuras}
\renewcommand{\listtablename}{Lista de Tabelas}
\renewcommand{\refname}{Referências}
\renewcommand{\indexname}{Índice}
\def\figurename{Fig.} %% Não é necessário ser trocado.
\renewcommand{\tablename}{TABELA}
\renewcommand{\partname}{Parte}
\renewcommand{\appendixname}{Apêndice}
\renewcommand{\abstractname}{Resumo}
% IEEE specific names
\renewcommand{\IEEEkeywordsname}{Palavras-chave}
\renewcommand{\IEEEproofname}{Prova}
%%%%%%%%%%%%%%%%%%%%%%%%%%%%%%%%%%%%%%%%%%%%%%%%%%%%%%%%%%%%%%%%%%%%%%%%%%%%
%%%%%%%%%%%%%%%%%%%%%%%%%%%%%%%%%%%%%%%%%%%%%%%%%%%%%%%%%%%%%%%%%%%%%%%%%%%%

%%%%%%%%%%%%%%%%%%%%%%%%%%%%%%%%%%%%%%%%%%%%%%%%%%%%%%%%%%%%%%%%%%%%%%%%%%%%
%*** NÃO ALTERE 
%%%%%%%%%%%%%%%%%%%%%%%%%%%%%%%%%%%%%%%%%%%%%%%%%%%%%%%%%%%%%%%%%%%%%%%%%%%%
%*** O TÍTULO DEVE SER DEFINIDO NO LOCAL APROPRIADO, NO INÍCIO DESSE ARQUIVO.
\title{~\\[2ex] \TITULO}
%%%%%%%%%%%%%%%%%%%%%%%%%%%%%%%%%%%%%%%%%%%%%%%%%%%%%%%%%%%%%%%%%%%%%%%%%%%%
%%%%%%%%%%%%%%%%%%%%%%%%%%%%%%%%%%%%%%%%%%%%%%%%%%%%%%%%%%%%%%%%%%%%%%%%%%%%

\author{
Wesley dos Reis Bezerra,~\IEEEmembership{PhD~Candidate,~PPGCC/UFSC},\\
Carlos Becker Westphall,~\IEEEmembership{Prof. Dr. PPGCC/UFSC},\\
\thanks{Autor correspondente: Wesley dos Reis Bezerra, \textit{wesleybez@gmail.com}}
}

%*** NÃO ALTERE 
%%%%%%%%%%%%%%%%%%%%%%%%%%%%%%%%%%%%%%%%%%%%%%%%%%%%%%%%%%%%%%%%%%%%%%%%%%%%
%*** SE NECESSÁRIO, ESSAS INFORMAÇÕES 
%*** SERÃO ATUALIZADAS PELOS EDITORES
%\markboth{\scriptsize \AUTORES\ / \RSIFSMA\ n. \NUM\ (\ANO) pp. \PP}%
%{\AUTORES: \TITULO}
%%%%%%%%%%%%%%%%%%%%%%%%%%%%%%%%%%%%%%%%%%%%%%%%%%%%%%%%%%%%%%%%%%%%%%%%%%%%
%%%%%%%%%%%%%%%%%%%%%%%%%%%%%%%%%%%%%%%%%%%%%%%%%%%%%%%%%%%%%%%%%%%%%%%%%%%%

\maketitle

\renewcommand{\abstractname}{Abstract}
\begin{abstract}
\boldmath
The few resources available on devices restricted in Internet of Things are an important issue when we think about security. In this perspective, our work proposes a agile systematic review literature on works involving the Internet of Things, authentication, and Fog Computing. As a result, related works, opportunities, and challenges found at these areas' intersections were brought, supporting other researchers and developers who work in these areas.
\end{abstract}
%%%%%%%%%%%%%%%%%%%%%%%%%%%%%%%%%%%%%%%%%%%%%%%%%%%%%%%%%%%%%%%%%%%%%%%%%%%%

%%%%%%%%%%%%%%%%%%%%%%%%%%%%%%%%%%%%%%%%%%%%%%%%%%%%%%%%%%%%%%%%%%%%%%%%%%%%
%*** PALAVRAS-CHAVE EM INGLÊS, SEPARADAS POR VÍRGULAS
%%%%%%%%%%%%%%%%%%%%%%%%%%%%%%%%%%%%%%%%%%%%%%%%%%%%%%%%%%%%%%%%%%%%%%%%%%%%
\renewcommand{\IEEEkeywordsname}{Index Terms}
%%%%%%%%%%%%%%%%%%%%%%%%%%%%%%%%%%%%%%%%%%%%%%%%%%%%%%%%%%%%%%%%%%%%%%%%%%%%

% \fi
\IEEEpeerreviewmaketitle

\section{Introduction}
\label{S:1}

Security is a challenge in several areas of computing, especially for the Internet of Things (IoT)\cite{zhang2019physical}. Specifically, when it has few resources\cite{deep2020survey,cao2019survey,yi2015fog}, such as the battery, processing, storage, and throughput; the use of security can be relegated. Thus, devices that have fewer resources tend to implement weaker security.

However, even with few resources, it is important to properly implement security in IoT\cite{ometov2018multi} systems $-$ in our case the implementation of device authentication. For that, it is necessary to know this specific area's trends, challenges, and opportunities. There was a lack of a preliminary study with the desired focus.

As a solution, this work aims to bring documentary resources that represent possible future developments in the area but also expose abysses that researchers/developers must avoid. This material was obtained from a agile systematic literature review (SLR) to answer the following questions.
\begin{itemize}
    \item what are the opportunities for IoT authentication in Fog Computing?
    \item what are the challenges for IoT authentication in Fog Computing?
    \item what are the main works related to the topic of IoT authentication at Fog Computing?
\end{itemize}

The work is organized as follows: in the second section, the materials and methods used were presented; then, in the third section, a agile systematic literature review was presented; in the next section, there is the quantitative analysis of the data obtained from the researched works; finally, the conclusion and future works were brought in the sixth section.

\begin{comment}

\begin{table}[h!]
    \centering
    \caption{Abbreviation list}
    \begin{tabular}{ll}
    \toprule
        IoT & Internet of Things  \\
        EBSE & Evidence Based Software Engineering\\
        CSV & Comma Separated Value\\
        BA & Bibliometric Analyses\\
        OTP & On-Time Password\\
        MQTT & Message Queuing Telemetry Transport\\
        SCADA & Supervisory Control and Data Acquisiton\\
        6LoWPAN & IPv6 over Low-Power Wireless Personal Area Network\\
        RPL & IPv6 Routing Protocol for Low Power and Lossy Networks\\
        CoAP & Constrained Application Protocol\\
        JWT & JSON Web Token\\
        UDP & User Datagram Protocol\\
        \bottomrule
    \end{tabular}
    
    \label{tab:my_label}
\end{table}
\end{comment}

\section{Materials and methods}
\label{S:3}

\begin{figure*}[h!]
    \centering
    \includegraphics[width=0.75\textwidth]{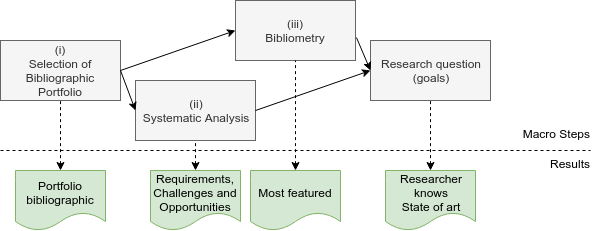}
    \caption{ProKnow-C Macro Steps (adapted from \cite{chaves2012gestao}) }
    \label{fig:proknowc_processs}
\end{figure*}

This study utilized the customization of the ProKnow-C \cite{chaves2012gestao} and EBSE \cite{kitchenham2004evidence} systematic review method focusing on the state of the art opportunities in the researched area, Figure \ref{fig:proknowc_processs}. The systematic review is a structural investigation which uses systematic procedures for searching, synthesis, and analysis \cite{machado2020analise} of the collected evidence. This methodology enabled the reduction of bias in surveying the bibliographic portfolio \cite{moher2009preferred}, obtaining quantitative data and a more focused, higher quality portfolio.

Significant tools contributed to greater quality and replication of the process. Mendeley \footnote{https://www.mendeley.com, which is an important research tool\cite{reiswig2010mendeley} among both students and other researchers\cite{zaugg2011mendeley}}, was used for managing the bibliographic portfolio, which allowed for the creation of folders to organize articles and store them while synchronized with the cloud. With respect to creating datasets, LibreOffice Calc \footnote{https://pt-br.libreoffice.org/ - an open-source software project for office automation with a strong community\cite{gamalielsson2014sustainability} } - was used to create the spreadsheets that documented the project and the datasets' generation in the CSV format to read later and create charts. Lastly, concerning the generation of charts, GNUPlot \footnote{http://www.gnuplot.info/ - a command line tool for the generation of charts \cite{williams20041} used by different IoT researchers\cite{wang2016improved,wang2018design,corona2019correlation} as well} - enabled the automation of charts in this work.

\section{Systematic Review of Literature}
\label{S:3.1}

\begin{table}[h!]
    \centering
    \caption{Research Areas - Results of the Analysis of the Related Areas and Publications - – on the left are listed the keywords followed by their results by year and increasingly by year}
    \begin{tabular}{cccc}
        \toprule
        Area              & 2018 & 2019 & 2020 \\
        \toprule
        IoT               & 83700 & 59300 & 31800  \\
        Authentication    & 42700 & 20400 & 21700 \\
        Fog Computing     & 16100 & 12100 & 4600 \\
        Message Protocol  & 3330  & 1820  & 520  \\
        \bottomrule
    \end{tabular}
    
    \label{tab:rsl_research_areas}
\end{table}

The ACM Digital Library, IEEE Xplorer, Scopus, ScienceDirect, and Scielo portals were selected. All chosen portals allow access to many publications in journals and conferences. Moreover, such portals permit free access to many publications through partnerships between universities and CAPES\footnote{https://www.periodicos.capes.gov.br/}.

Concerning the engineering process of query-string, four main areas were chosen. The selected areas were IoT, authentication, fog computing, and message protocols. The results can be seen in Figure \ref{tab:rsl_research_areas}, which offers an overview of the last three years of publications in each area. A three-year time window (2018-2020) was chosen due to the present study only seeking new works and trends in the mentioned areas. For the initial exploration of the number of publications obtained, the Google Scholar \footnote{http://scholar.google.com} tool was used, without the inclusion of patents and citations.

\begin{figure}[h!]
    \centering
    \includegraphics[width=0.45\textwidth]{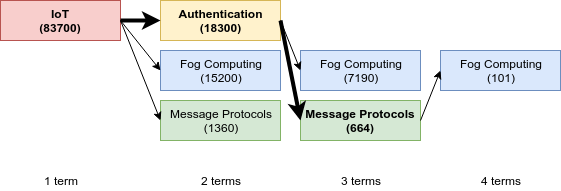}
    \caption{Query Terms Tree - Results of the Analysis of the Related Publications – from the left, the search terms are combined. This combination is made with up to four terms in the rightmost column and follows the flow of each connection through the directional arrows}
    \label{fig:query_tree}
\end{figure}

During the analysis of areas,  a query terms tree was created \ref{fig:query_tree}. In this tree, the paths taken to combine the areas forming the consultation questions can be seen. In highlights are the words used in the chosen query-string. It is evident that even having a consultation question with four terms, this did not present satisfactory results when submitted to the consultation portals. When submitted to such portals, it brought few or no results. Correspondingly, we opted for the query-string of three terms: (IoT AND authentication AND "message protocols").

After submitting the query-string to the portals, it can be noticed that the portal with the highest number of results was the Digital Library ACM with eight publications, followed by Scopus with four publications. In the total sum of the results among all  portals listed, 13 publications appeared in the last three years.

Subsequently, publications that did not meet the established criteria were removed from the articles' initial list. The inclusion and exclusion criteria are significant and directly influence the selected publications' quality \cite{kitchenham2009systematic}. It can be seen that the criteria has been divided into three inclusion factors and four exclusion factors. As the time window was used when consulting portals, it was unnecessary to include a criterion referencing maximum time for evaluated publications.

\begin{comment}

\begin{table}[h!]
    \centering
    \caption{Inclusion and exclusion criteria used }
    \begin{tabular}{cl}
    \toprule
    Inclusion     &  \\
    \midrule
    I01     & Complete articles from conference or periodicals; \\
    I02     & English only language articles; \\
    I03     & Full-text access available.\\
    \midrule
    Exclusion & \\
    \midrule
    E01     & Research patents and reports; \\
    E02     & Non-scientific publication; \\
    E03     & For 2018 and 2019 publications, less than five citations;\\
    E04     & Duplicate publications.\\
    \bottomrule   
    \end{tabular}
    \label{tab:criterio_inc_exc}
\end{table}

\end{comment}

After defining the criteria, there was the portfolio selection. The process began with 13 publications, two of which were eliminated because they were non-scientific publications, and two others were eliminated because they were duplicate publications. In the title reading phase, nine publications were analyzed, of which one publication was eliminated due to not being a title adhering to the scope of the research. In the reading phase of abstracts, two publications were eliminated, leaving only six publications in the list of articles. In the last phase, the complete reading phase, no publication was eliminated, leaving six final portfolio publications. 

\begin{comment}

\end{comment}

The final portfolio, Table \ref{tab:tab_portifolio}, is composed of six publications. Even though this is not an expressive number, the selected publications brought a large and diverse amount of information about the research topic. It is also observed that two of the six publications are from journals and bring a comprehensive view on the state-of-the-art and challenges related to the theme. The other publications contribute mainly to what types of solutions are being given to  authentication with IoT message protocols.

\begin{table}[h!]
\centering
\caption{Bibliographic portfolio - the first column contains an index of the publication within the database used to support the systematic review and the reference; the second column contains the title of the evaluated publication}
\Rotatebox{90}{
    \begin{tabular}{rll}
    \toprule
    \# & Document   & Publication  \\ 
    \toprule
    a1 & \cite{buccafurri2019blockchain}	& \makecell[l]{A Blockchain-Based OTP-Authentication Scheme for Constrained IoT Devices Using MQTT}\\
    a3 & \cite{brandao2020blockchain} & \makecell[l]{A Blockchain-Based Protocol for Message Exchange in a ICS Network: Student Research Abstract}\\
    a4 & \cite{mahmoud2019security}	& \makecell[l]{Security for Internet of Things: A State of the Art on Existing Protocols and Open Research Issues}\\
    a6 &\cite{usman2019survey} & \makecell[l]{A Survey on Representation Learning Efforts in Cybersecurity Domain}\\
    a9 & \cite{iyer2018implementation} & \makecell[l]{Implementation and Evaluation of Lightweight Ciphers in MQTT Environment}\\
    a11 & \cite{bhawiyuga2017architectural} & \makecell[l]{Architectural design of token based authentication of MQTT protocol in constrained IoT device}\\
    \bottomrule
    \end{tabular}
    }
    
    \label{tab:tab_portifolio}
\end{table}

\subsection{Quantitative Data Analysis}
\label{S:3.2}

In bibliometric analyses (BA), statistical metrics found on selected documents are expressed through a pictograph \cite{dabic2020pathways}. A more quantitative analysis of the research area is possible from the collected variables. Some expected results understand the main events and journals in the area, as well as the leading and most relevant authors, among other variables. It is of significant importance to know which publications are most influential by comparing the number of citations.

\begin{figure}[h!]
    \centering
    \subfigure[Publications by year]{
        \includegraphics[width=0.45\textwidth]{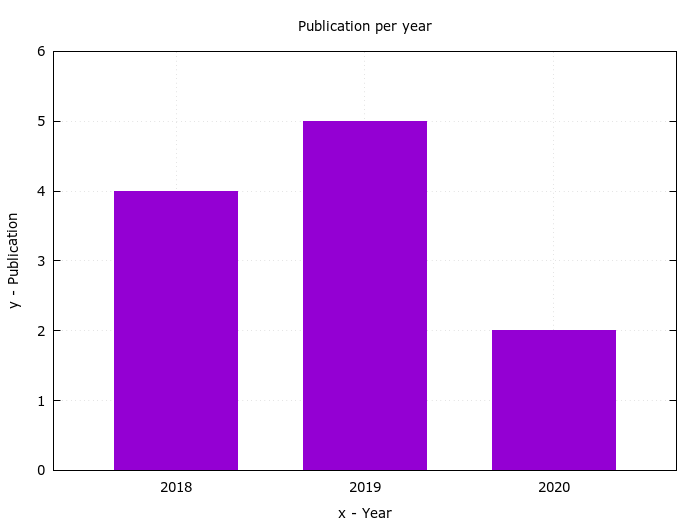}
    }
    \hfill
    \subfigure[Citations per publication]{
        \includegraphics[width=0.45\textwidth]{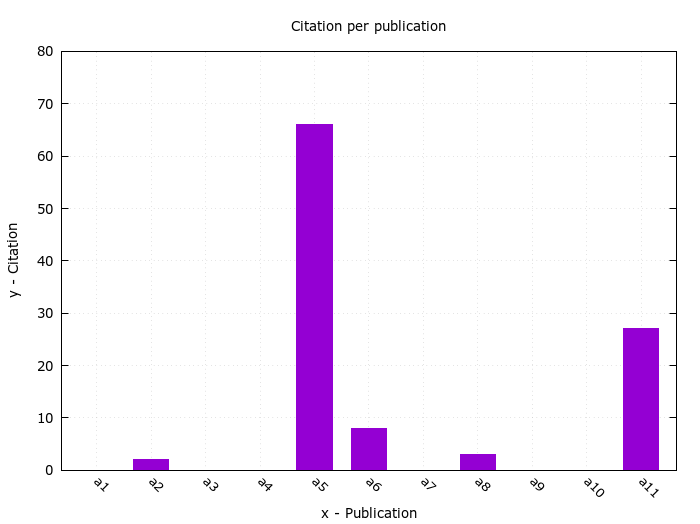}
    }
    \caption{Publication Data - bibliometric data on published studies. As can be seen on the left, ranging from 2018 to 2020, is the distribution of the number of publications per year. On the right, the distribution of citations per publication throughout the study period}
    \label{fig:publication_data}
\end{figure}

\begin{figure}[h!]
    \subfigure[h-index]{
        \includegraphics[width=0.45\textwidth]{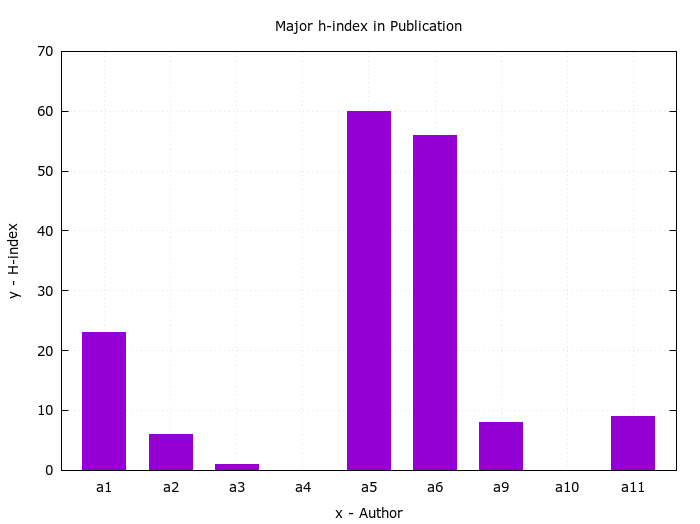}
    }
    \hfill
    \subfigure[i10]{
        \includegraphics[width=0.45\textwidth]{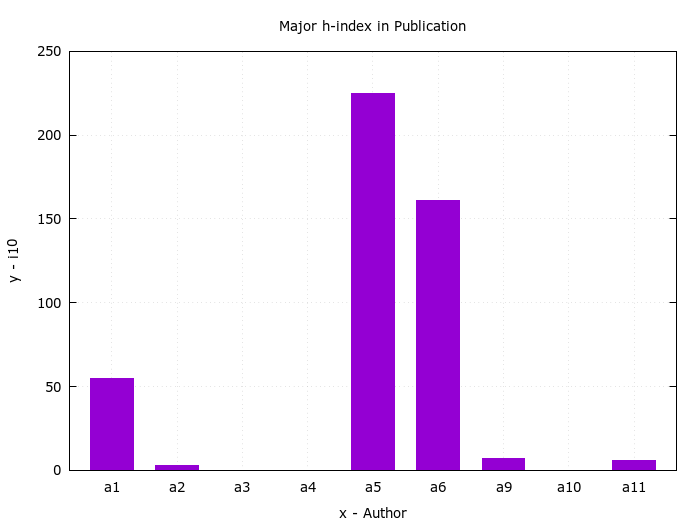}
    }
    \caption{Highest Indexes among the Authors of Each Publication - in these bibliometric graphs, an analysis of the authors participating in the portfolio is described. In these graphs, the authors are associated with their publications and share the same index as their publications. On the left, there is a description of the largest h-index of the publication author. On the right, is listed the highest i10 of the author of each publication}
    \label{fig:major_indexes}
\end{figure}

\begin{figure}[h!]
    \subfigure[Publications by Conference/Journal]{
        \includegraphics[width=0.45\textwidth]{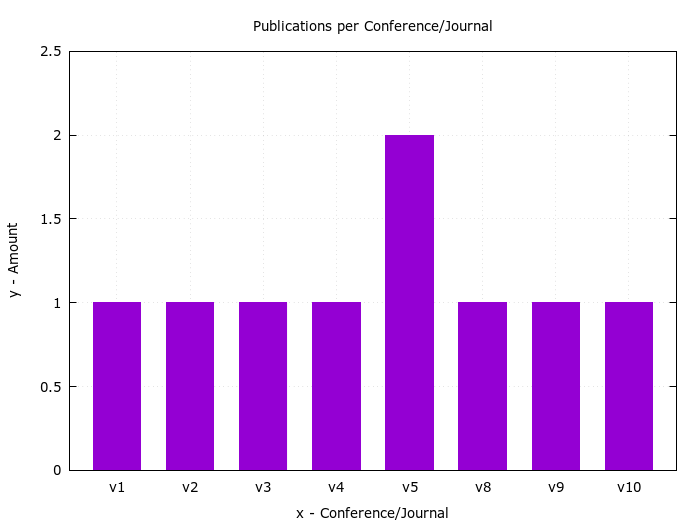}
    }
    \hfill
    \subfigure[Journal Impact Factor]{
        \includegraphics[width=0.45\textwidth]{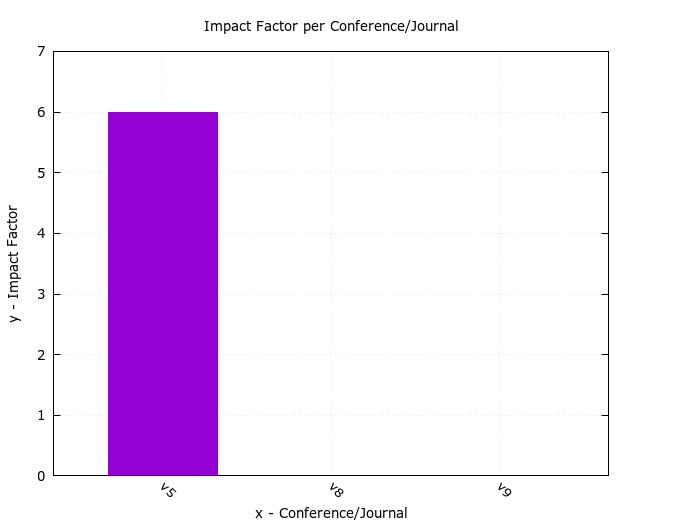}
    }
    \caption{Conferences and Journal Data - These graphs represent a description of the sources utilized in this systematic review. On the left, is a description of the number of publications per source. Presented on the right, is a description of the impact factors of the journals which are part of the research}
    \label{fig:vehicle}
\end{figure}

Three dimensions were chosen to be analyzed bibliometrically: publications, authors, and sources. The dimensions of the publications are shown in Figures \ref{fig:publication_data}-(a) and \ref{fig:publication_data}-(b) which show the evolution of the works published during the selected period and an analysis of citations per study. Regarding the authors' dimension, Figures \ref{fig:major_indexes}-(a) and \ref{fig:major_indexes}-(b), weigh the significance of the authors involved in the publications and suggest researchers to whom research should be monitored in order to remain up-to-date in the area. Lastly, the dimension of the data sources are visualized through Figures \ref{fig:vehicle}-(a) and \ref{fig:vehicle}-(b) which enable the reader to identify the main sources of research within the proposed analysis, their impact factors, and the quantity of publications. It can be concluded that the three dimensions permit the identification of the most relevant studies, the authors which should be followed and the sources to aim for in order to obtain publications.

This work brings a BA of the entire researched area and not just the portfolio list. This analysis was a project choice due to the reduced number of publications in the final portfolio. The authors considered that even if there were no complete adherence to all articles found, the articles still described the area of interest in the last three years, the time window established at the time of consultation.

The variables collected from the documents helped  inferences to be drawn from the publications. The following variables were collected: publications per year (\ref{eq:pub_year}), highest h-index per publication (\ref{eq:major_hindex}), highest i10 per publication (\ref{eq:major_i10}), publications per conference/journal, the journal impact factor (when available), h-index of each conference (when available) and citations per publication. 

\begin{equation}
    publicationYear_i = \sum\left ( count(publication_i)) \right )
    \label{eq:pub_year}
\end{equation}

\begin{equation}
    majorHindex_i = Major\left(hindex(author_j,publication_i))\right)
    \label{eq:major_hindex}
\end{equation}

\begin{equation}
    majorI10_i = Major\left(i10(author_j,publication_i))\right)
    \label{eq:major_i10}
\end{equation}

As is evident in Figure \ref{fig:publication_data}-(a), the publications' distribution accumulated (\ref{eq:pub_year}) more results in the first two years. The year 2020 had fewer results, partly because publications were ongoing. Additionally, some events delayed their achievements and resulted in the annals' publication.

The number of citations is an essential factor for assessing the impact of a job on the academic community. Most citations were published by a5 \cite{siow2018analytics} with 66 citations, followed by a11 \cite{bhawiyuga2017architectural} with 27 citations. It is observed that the most recent studies, although relevant, have not yet had the opportunity to generate citations or the citations generated have not been indexed yet.

Regarding the authors' quality assessment, two indexes were chosen, the h-index (\ref{eq:major_hindex}) and the i10(\ref{eq:major_i10}). For the i10 index, Figure \ref{fig:major_indexes}-(b), the highest value is associated with the researcher Wendy Hall of the publication a5 \cite{siow2018analytics} with 225 points. As for the H index (\textit{h-index}), Figure \ref{fig:major_indexes}-(a), the same author leads with 60 points, followed by Jinjun Chen from a6 \cite{usman2019survey} with 56 points.

Publications were found in 11 different publishing vehicles, Figure \ref{fig:vehicle}-(a). Of these vehicles found, only two were journals, being the other conferences' totality. It is quite noticeable that even with only two journals in the list of publications, the publications brought forth important material for discussion (a5, a6, and a10). 
\begin{comment}

\begin{table*}[h!]
\centering
\caption{Conferences and Journal of publications found}
\adjustbox{max width=\textwidth}{
    
    \begin{tabular}{rl}
    \toprule
\#	&Vehicle \\
\toprule
v1	&\makecell[l]{Proceedings of the 2019 3rd International Symposium on Computer Science and Intelligent Control}\\
v2	&\makecell[l]{Proceedings of the Ninth International Symposium on Information and Communication Technology}\\
v3	&\makecell[l]{Proceedings of the 35th Annual ACM Symposium on Applied Computing}\\
v4	&\makecell[l]{Proceedings of the 9th International Conference on Information Systems and Technologies}\\
v5	&\makecell[l]{ACM Computer Surveys}\\
v8	&\makecell[l]{2018 International Conference on Electrical, Electronics, Communication, Computer, and Optimization Techniques (ICEECCOT)}\\
v9	&\makecell[l]{International Journal of Advanced Science and Technology}\\
v10	&\makecell[l]{Proceeding of 2017 11th International Conference on Telecommunication Systems Services and Applications, TSSA 2017}\\
\bottomrule    
    \end{tabular}
    }
    \label{tab:vehicle}
\end{table*}
\end{comment}

The journal with the most significant impact factor is ACM Computer Surveys, Figure \ref{fig:vehicle}-(b), with an index of 6.131. From this journal, two articles were analyzed (a5 and a6). Such a journal is renowned for bringing quality and extensive content on several computing areas, serving as a reliable source in the search for evidence to support projects and research choices in the area. This result reinforced the statements previously made.

It can be seen through bibliometrics that the theme has grown in many publications over recent years. It is also possible to note relevant researchers' publications in the area. Such researchers have published in relevant journals and conferences. 

\section{Discussion and Results}
\label{S:3.3}

For thematic review, the challenge lens and the opportunity lens were chosen. The challenge lens aims to list the challenges encountered by each article in the portfolio, highlighting the challenges of the chosen area. Conversely, the opportunities lens enables the researcher to analyze which solutions are used and research opportunities within the area. Together, the two lenses allowed us to see the research characteristics carried out on Authentication in IoT for message protocols.

In Buccafurri and Romolo \cite{buccafurri2019blockchain}, the authors proposed an authentication integrating OTP and blockchain (smart contracts in Ethereum) to solve the fragile existing problem authentication mechanisms. The authors used MQTT as a message protocol in their experiment. They noted the challenges in which restricted devices often do not support the implementation of security mechanisms and fragility of existent ones. They commented on the need for future work to implement and validate its proposal and security analysis.

Brandão \cite{brandao2020blockchain} also introduced a solution using blockchain. Its application area controls industrial systems, specifically in the SCADA system (Supervisory Control and Data Acquisition). Blockchain was used to resolve security and data storage issues on devices. The author presented challenges regarding the number of devices participating; the greater the number of devices, the more expensive the synchronization process.

Mahmoud and Aouag \cite{mahmoud2019security} brought a state-of-the-art view on protocols and research opportunities within the scope of security for IoT. They addressed security in 6LoWPAN, RPL, and CoAP, thus bringing a multi-layered view of the protocol stack for IoT systems. This work presented the Sybil attack and the fabrication of identities as a relevant threat to the systems mentioned earlier. Among the concerning issues in the security area for RPL were blackhole attacks, attack wormholes, and version number attacks. They also showed blockchain as a trend participant in IoT security solutions. Future studies will bring more depth about the attacks, evaluating the solutions with blockchain and how the attacks on the RPL affect the devices' consumption of resources.

Usman \textit{et al.} \cite{usman2019survey} put forward a study on the representation of learning in the security domain. The authors proposed a set of machine-learning and threat detection techniques. It also analysed the platforms which offer machine-learning in the cloud, which allowed the researcher to have support in choosing the platform used. The most used security data sets on security are also presented. Limitations and challenges were presented on the exposed data sets and techniques.

Iyer \textit{et al.} \cite{iyer2018implementation} implemented and evaluated lightweight algorithms for data security in the MQTT protocol. This research established the premise that a lightweight hash function and lightweight symmetric encryption were good combinations for MQTT security with low resource consumption. It also focused on the payload and used SPONGENT\footnote{a lightweight hash function based on a wide PRESENT-type permutation\cite{bogdanov2011spongent}} as a hash solution to bring confidentiality, integrity, and authenticity to the information transferred.

Bhawiyuga, Data, and Warda \cite{bhawiyuga2017architectural} altered the standard architecture of publish/subscribe applications, inserting the authentication token server's figure. The server is responsible for authenticating and releasing the authentication token to enable communication between the parties. This solution used the JWT\footnote{JSON Web Token - the security JSON token which enables cross security domain sharing of identity and security information\cite{jones2015json}} server to distribute the tokens. They also presented a performance test with the proposed new architecture.

Several areas are covered in the bibliographic portfolio. An overview of the challenges was listed in each work, and each set of authors also commented on the proposed solution. The systematic review raised the necessary data for the list to be commented in the next section of this paper, the section on Opportunities and Challenges.

\subsection{Opportunities and Challenges}
\label{S:3.4}

This section presents the main results obtained from the SLR of our work. The results obtained from the bibliographic portfolio are synthesized and discussed. Our work puts forward a list of opportunities and challenges found in SRL. Two lists are presented: initially, the opportunities found are discussed, and subsequently, the challenges listed by some selected works are commented on. First, the \textbf{opportunities}:

\begin{itemize}
    \item wide use of the MQTT protocol \cite{buccafurri2019blockchain, iyer2018implementation, bhawiyuga2017architectural};
    \item use of \textit{blockchain} \cite{buccafurri2019blockchain,brandao2020blockchain,mahmoud2019security};
    \item use of lightweight approaches \cite{iyer2018implementation,bhawiyuga2017architectural};
\end{itemize}

MQTT message protocol appeared in half of the works in the bibliographic portfolio. This old protocol is a messaging protocol widely used in production and widely accepted in academia. Only one study presented CoAP as a messaging protocol solution. However, this is an option with low impact for networks of restricted devices. This is due to the fact that such a protocol travels over UDP.

Technologies associated with blockchain have been recurrent in the works presented. This technology is gaining acceptance as a security solution in smart transactions and contracts. Its use appears in several proposed solutions for IoT security, being a topic of relevance within security research and message protocols.

A constant in the studies presented is the need for a light approach to such devices. The studies which implemented security solutions \cite{iyer2018implementation, bhawiyuga2017architectural} reached a consensus on preserving computational resources. Another important consensus was that the security which exists in the protocols today is not adequate to the security level required for real IoT systems.

Just as the opportunities were exposed, we also present the \textbf{challenges} listed by our portfolio. As per the main challenges, the following can be listed:
\begin{itemize}
    \item need for further validation and evaluation \cite{buccafurri2019blockchain};
    \item sync between the most devices when using blockchain \cite{brandao2020blockchain};
    \item analysis of resource consumption and ways to use blockchain \cite{mahmoud2019security};
    \item improvements in datasets and machine-learning techniques used for cybersecurity \cite{usman2019survey}.
\end{itemize}

Due to the wide variety of IoT application scenarios, the evaluation and validation of solutions present themselves as challenges. It is impossible to validate a solution which performs optimally in all situations. Thus, reassessing the solutions in each scenario is the best solution.

The increase in the number of solutions involving blockchain is unquestionable. However, it is necessary to evaluate its adoption in each area. The issue of synchronization between the parties can cause a delay in the development of tasks and make the system unfeasible. Some areas have a better supply of resources and access to more robust hardware, but this is not the reality in all areas of the IoT. An approach that fits resource consumption needs would be the most favorable solution in these areas.

Improved and more specific datasets would help in training solutions which use machine learning. The improvement and optimization of the techniques used would also provide better means for research in the IoT area. Algorithms which consider devices' limitations or avoid such limitations with a different approach.

Many other challenges exist in the literature to be listed along with these. Our work provides an overview of recent articles. Further research must be carried out for the researcher's challenges in a specific IoT area. As cited, changes in areas can change solutions' performance or even make them inadequate \cite{dos2020ambiente}.

\section{Conclusion}
\label{S:6}

The SLR successfully brought the desired answers, presenting the opportunities and challenges for the authentication of IoT devices in Fog Computing. A bibliographic portfolio, a quantitative analysis of the documents, and a discussion of the works found were also raised. As a result, this work presented some research material that can be reused in future technology research and development.

In future works, we propose a more comprehensive search and meta-analysis of the \textit{corpus} of documents found. Also, develop work for continuous monitoring of opportunities so that they are better used; and the challenges for its mitigation.

\ifCLASSOPTIONcaptionsoff
  \newpage
\fi



%\bibliographystyle{IEEEtran}
%\bibliography{bibs/geral.bib,bibs/desafios.bib,bibs/paradigmas.bib}
%\begin{thebibliography}{1}
%\end{thebibliography}

\printbibliography

@inproceedings{yi2015fog,
  title={Fog computing: Platform and applications},
  author={Yi, Shanhe and Hao, Zijiang and Qin, Zhengrui and Li, Qun},
  booktitle={2015 Third IEEE Workshop on Hot Topics in Web Systems and Technologies (HotWeb)},
  pages={73--78},
  year={2015},
  organization={IEEE}
}

@inproceedings{dos2020ambiente,
  title={Ambiente de experimenta{\c{c}}{\~a}o para avalia{\c{c}}{\~a}o protocolos de mensagem para IoT na Fog},
  author={dos Reis Bezerra, Wesley and Westphall, Carlos Becker},
  booktitle={Anais do Workshop de Pesquisa Experimental da Internet do Futuro},
  pages={1--6},
  year={2020},
  organization={SBC}
}

@article{reiswig2010mendeley,
  title={Mendeley},
  author={Reiswig, Jennifer},
  journal={Journal of the Medical Library Association: JMLA},
  volume={98},
  number={2},
  pages={193},
  year={2010},
  publisher={Medical Library Association}
}

@article{zaugg2011mendeley,
  title={Mendeley: Creating communities of scholarly inquiry through research collaboration},
  author={Zaugg, Holt and West, Richard E and Tateishi, Isaku and Randall, Daniel L},
  journal={TechTrends},
  volume={55},
  number={1},
  pages={32--36},
  year={2011},
  publisher={Springer}
}

@article{gamalielsson2014sustainability,
  title={Sustainability of Open Source software communities beyond a fork: How and why has the LibreOffice project evolved?},
  author={Gamalielsson, Jonas and Lundell, Bj{\"o}rn},
  journal={Journal of Systems and Software},
  volume={89},
  pages={128--145},
  year={2014},
  publisher={Elsevier}
}

@article{williams20041,
  title={1 gnuplot},
  author={Williams, Thomas and Kelley, Colin and Lang, Russell and Kotz, Dave and Campbell, John},
  year={2004}
}

@inproceedings{wang2016improved,
  title={An Improved Cluster Routing Structure of IOT},
  author={Wang, Zhihui and Wu, Ruokun and Sa, Qila and Li, Jinlin and Fan, Yuejiao and Xu, Wenbo and Zhao, Yiqun},
  booktitle={2016 International Conference on Communications, Information Management and Network Security},
  pages={326--328},
  year={2016},
  organization={Atlantis Press}
}

@inproceedings{wang2018design,
  title={Design and implementation of NS3-based simulation system of LEO satellite constellation for IoTs},
  author={Wang, Zheng and Cui, Gaofeng and Li, Pengxu and Wang, Weidong and Zhang, Yinghai},
  booktitle={2018 IEEE 4th international conference on computer and communications (ICCC)},
  pages={806--810},
  year={2018},
  organization={IEEE}
}

@inproceedings{corona2019correlation,
  title={Correlation Study between Photovoltaic Power Output and Environmental Variables Using an Embedded IoT System},
  author={Corona-Ventura, Jason Y and Lobato-Nostroza, Oscar and Ch{\'a}vez-Campos, Gerardo M and Lara-Hern{\'a}ndez, Rafael and Chiariada-Masseli, Yvo M and T{\'e}llez-Anguiano, Adriana C and Fraga-Aguilar, Miguelangel},
  booktitle={2019 IEEE International Autumn Meeting on Power, Electronics and Computing (ROPEC)},
  pages={1--6},
  year={2019},
  organization={IEEE}
}

@article{deep2020survey,
  title={A survey of security and privacy issues in the Internet of Things from the layered context},
  author={Deep, Samundra and Zheng, Xi and Jolfaei, Alireza and Yu, Dongjin and Ostovari, Pouya and Kashif Bashir, Ali},
  journal={Transactions on Emerging Telecommunications Technologies},
  pages={e3935},
  year={2020},
  publisher={Wiley Online Library}
}

@article{dabic2020pathways,
  title={Pathways of SME internationalization: a bibliometric and systematic review},
  author={Dabi{\'c}, Marina and Maley, Jane and Dana, Leo-Paul and Novak, Ivan and Pellegrini, Massimiliano M and Caputo, Andrea},
  journal={Small Business Economics},
  volume={55},
  number={3},
  pages={705--725},
  year={2020},
  publisher={Springer}
}

@inproceedings{bogdanov2011spongent,
  title={SPONGENT: A lightweight hash function},
  author={Bogdanov, Andrey and Kne{\v{z}}evi{\'c}, Miroslav and Leander, Gregor and Toz, Deniz and Var{\i}c{\i}, Kerem and Verbauwhede, Ingrid},
  booktitle={International Workshop on Cryptographic Hardware and Embedded Systems},
  pages={312--325},
  year={2011},
  organization={Springer}
}

@article{jones2015json,
  title={JSON Web Token (JWT) profile for OAuth 2.0 client authentication and authorization Grants},
  author={Jones, Michael and Campbell, Brain and Mortimore, Chuck},
  journal={May-2015.[Online]. Available: https://tools. ietf. org/html/rfc7523},
  year={2015}
}

@article{zhang2019physical,
  title={Physical layer security for the Internet of Things: Authentication and key generation},
  author={Zhang, Junqing and Rajendran, Sekhar and Sun, Zhi and Woods, Roger and Hanzo, Lajos},
  journal={IEEE Wireless Communications},
  volume={26},
  number={5},
  pages={92--98},
  year={2019},
  publisher={IEEE}
}

@inproceedings{buccafurri2019blockchain,
  title={A Blockchain-Based OTP-Authentication Scheme for Constrainded IoT Devices Using MQTT},
  author={Buccafurri, Francesco and Romolo, Celeste},
  booktitle={Proceedings of the 2019 3rd International Symposium on Computer Science and Intelligent Control},
  pages={1--5},
  year={2019}
}

@inproceedings{brandao2020blockchain,
  title={A blockchain-based protocol for message exchange in a ICS network: student research abstract},
  author={Brand{\~a}o, Ricardo},
  booktitle={Proceedings of the 35th Annual ACM Symposium on Applied Computing},
  pages={357--360},
  year={2020}
}

@inproceedings{mahmoud2019security,
  title={Security for Internet of Things: A State of the Art on existing Protocols and Open Research issues},
  author={Mahmoud, Chaira and Aouag, Sofiane},
  booktitle={Proceedings of the 9th International Conference on Information Systems and Technologies},
  pages={1--6},
  year={2019}
}

@inproceedings{iyer2018implementation,
  title={Implementation and Evaluation of Lightweight Ciphers in MQTT Environment},
  author={Iyer, Shweta and Bansod, GV and Naidu, Praveen and Garg, Shefali},
  booktitle={2018 International Conference on Electrical, Electronics, Communication, Computer, and Optimization Techniques (ICEECCOT)},
  pages={276--281},
  year={2018},
  organization={IEEE}
}

@inproceedings{bhawiyuga2017architectural,
  title={Architectural design of token based authentication of MQTT protocol in constrained IoT device},
  author={Bhawiyuga, Adhitya and Data, Mahendra and Warda, Andri},
  booktitle={2017 11th International Conference on Telecommunication Systems Services and Applications (TSSA)},
  pages={1--4},
  year={2017},
  organization={IEEE}
}

@article{siow2018analytics,
  title={Analytics for the internet of things: A survey},
  author={Siow, Eugene and Tiropanis, Thanassis and Hall, Wendy},
  journal={ACM Computing Surveys (CSUR)},
  volume={51},
  number={4},
  pages={1--36},
  year={2018},
  publisher={ACM New York, NY, USA}
}

@article{usman2019survey,
  title={A survey on representation learning efforts in cybersecurity domain},
  author={Usman, Muhammad and Jan, Mian Ahmad and He, Xiangjian and Chen, Jinjun},
  journal={ACM Computing Surveys (CSUR)},
  volume={52},
  number={6},
  pages={1--28},
  year={2019},
  publisher={ACM New York, NY, USA}
}

@article{moher2009preferred,
  title={Preferred reporting items for systematic reviews and meta-analyses: the PRISMA statement},
  author={Moher, David and Liberati, Alessandro and Tetzlaff, Jennifer and Altman, Douglas G},
  journal={Annals of internal medicine},
  volume={151},
  number={4},
  pages={264--269},
  year={2009},
  publisher={Am Coll Physicians}
}

@article{chaves2012gestao,
  title={Gest{\~a}o do processo decis{\'o}rio: mapeamento ao tema conforme as delimita{\c{c}}{\~o}es postas pelos pesquisadores},
  author={Chaves, Leonardo Corr{\^e}a and Ensslin, Leonardo and Ensslin, Sandra Rolim and Petri, Sergio Murilo and Da Rosa, Fabr{\'\i}cia Silva},
  journal={Revista Eletr{\^o}nica de Estrat{\'e}gia \& Neg{\'o}cios},
  volume={5},
  number={3},
  pages={3--27},
  year={2012}
}

@article{kitchenham2009systematic,
  title={Systematic literature reviews in software engineering--a systematic literature review},
  author={Kitchenham, Barbara and Brereton, O Pearl and Budgen, David and Turner, Mark and Bailey, John and Linkman, Stephen},
  journal={Information and software technology},
  volume={51},
  number={1},
  pages={7--15},
  year={2009},
  publisher={Elsevier}
}

@inproceedings{kitchenham2004evidence,
  title={Evidence-based software engineering},
  author={Kitchenham, Barbara A and Dyba, Tore and Jorgensen, Magne},
  booktitle={Proceedings. 26th International Conference on Software Engineering},
  pages={273--281},
  year={2004},
  organization={IEEE}
}

@article{machado2020analise,
  title={Uma an{\'a}lise sistem{\'a}tica da literatura acerca dos m{\'e}todos de usabilidade aplic{\'a}veis a dispositivos m{\'o}veis},
  author={Machado, Lais and Vergara, Lizandra Garcia Lupi},
  journal={Gepros: Gest{\~a}o da Produ{\c{c}}{\~a}o, Opera{\c{c}}{\~o}es e Sistemas},
  volume={15},
  number={1},
  pages={42},
  year={2020},
  publisher={Universidade Estadual Paulista-UNESP Bauru, Depto de Engenharia de Produ{\c{c}}{\~a}o}
}

@article{ometov2018multi,
  title={Multi-factor authentication: A survey},
  author={Ometov, Aleksandr and Bezzateev, Sergey and M{\"a}kitalo, Niko and Andreev, Sergey and Mikkonen, Tommi and Koucheryavy, Yevgeni},
  journal={Cryptography},
  volume={2},
  number={1},
  pages={1},
  year={2018},
  publisher={Multidisciplinary Digital Publishing Institute}
}

@article{cao2019survey,
  title={A Survey on Security Aspects for 3GPP 5G Networks},
  author={Cao, Jin and Ma, Maode and Li, Hui and Ma, Ruhui and Sun, Yunqing and Yu, Pu and Xiong, Lihui},
  journal={IEEE Communications Surveys \& Tutorials},
  year={2019},
  publisher={IEEE}
}

\end{document}